\documentclass[final,useAMS]{mn2e}
\usepackage{graphicx}
\usepackage{subfigure}
\font\twelvei = cmmi10 scaled\magstep1
       \font\teni = cmmi10 
\font\mbf = cmmib10 scaled\magstep1
       \font\mbfs = cmmib10 \font\mbfss = cmmib10 scaled 833
\font\msybf = cmbsy10 scaled\magstep1
       \font\msybfs = cmbsy10 \font\msybfss = cmbsy10 scaled 833
       \def\mit{\fam1 }
       \def\bmit{\fam9 }
\def\lsim{\raise0.3ex\hbox{$<$}\kern-0.75em{\lower0.65ex\hbox{$\sim$}}}
\def\gsim{\raise0.3ex\hbox{$>$}\kern-0.75em{\lower0.65ex\hbox{$\sim$}}}

\title[Numerical Simulation of Thick Accretion Flow]
{Numerical Simulation of Vertical Oscillations in an Axisymmetric Thick Accretion Flow around a Black Hole}
\author[Arnab Deb, Kinsuk Giri and Sandip K. Chakrabarti]
{Arnab Deb\thanks{arnab12a@bose.res.in}$^{1}$, Kinsuk Giri\thanks{kinsuk@nitttrkol.ac.in}$^{2}$, Sandip K. Chakrabarti
\thanks{chakraba@bose.res.in}$^{1,3}$\\
$^{1}$ S. N. Bose National Centre for Basic Sciences, Block -JD,  Sector -3, Salt Lake, Kolkata 700098, India\\
$^{2}$ National Institute of Technical Teachers Training and Research, Block-FC, Sector - 3, Salt Lake, Kolkata-700106, India \\
$^{3}$ Indian Centre for Space Physics, Chalantika 43, Garia Station Road, Kolkata-700084, India}

\begin{document}

\date{}


\maketitle

\label{firstpage}

\begin{abstract}

We study time evolution of rotating, axisymmetric, two dimensional inviscid accretion flows around black holes using  
a grid based finite difference method. We do not use reflection symmetry on the equatorial plane in order to inspect 
if the disk along with the centrifugal barrier oscillated vertically. In the inviscid limit, we find that the CENtrifugal pressure supported 
BOundary Layer (CENBOL) is oscillating vertically, more so, when the specific angular momentum is higher. As a result, 
the rate of outflow produced from the CENBOL, also oscillates. Indeed, the outflow rates in the upper half and the lower half are 
found to be anti-correlated. We repeat the exercise for a series of specific angular momentum $\lambda$ of the flow in order  
to demonstrate effects of the centrifugal force on this interesting behaviour. We find that, as predicted in   
theoretical models of disks in vertical equilibrium, the CENBOL is produced only when the centrifugal force is 
significant and more specifically, when $\lambda > 1.5$. Outflow rate itself is found to increase with $\lambda$
as well and so is the oscillation amplitude. The cause of oscillation appears to be due to the interaction among the back flow
from the centrifugal barrier, the outflowing winds and the inflow. For low angular momentum, the back flow as well as
the oscillation are missing. To our knowledge, this is the first time that such an oscillating 
solution is found with an well-tested grid based finite difference code and such a solution could be
yet another reason of why Quasi-Periodic Oscillations should be observed in black hole candidates which 
are accreting low angular momentum transonic flows.

\end{abstract}


\section{Introduction}

In astrophysical context, the process in which diffuse gas or matter is accumulated around a compact 
object under an influence of gravity is called accretion. The importance of accretion as the source 
of steady or unsteady emission of radiation was first widely recognized in explaining 
observations of binary systems, especially X-ray binaries. It is very important to understand 
hydrodynamic properties of matter in the vicinity of a black hole 
as the emitted radiation mainly depends on the density, velocity and temperature at each flow element at each instant. 
Unique inner boundary condition which enables all infalling matter to cross the event horizon with the speed of light 
$c$ (e.g., Chakrabarti, 1996) also makes all the accretion flows into black holes to be transonic in nature forcing 
them to have the innermost sonic point in  regions of strong gravity, i.e., just outside of the event horizon. 
Due to centrifugal pressure, inflowing 
matter slows down and piles up closer to the black hole forming a torus like structure at the inner part of the disc. 
This torus like structure formed between the centrifugally supported shock and the innermost sonic point outside the
horizon is widely known as the Centrifugal pressure supported BOundary Layer or simply CENBOL 
(Chakrabarti, 1989, hereafter C89; 1999; Molteni, Lanzafame \& Chakrabarti, 1994; MLC94). 
In the limit of no radial velocity, the CENBOL would have been termed as a thick accretion 
disk (Paczy\'nski \& Wiita, 1980) as shown by MLC94.

Earlier, a large number of numerical simulations of inviscid accretion flows around black holes, 
have been presented by various research groups (e.g., Hawley, Wilson \& Smarr 1984; 
Eggum, Coroniti, \& Katz 1985; Chakrabarti \& Molteni, 1993; MLC94; Ryu, Brown, Ostriker \& Loeb 1995;
Molteni, Sponholz \& Chakrabarti 1996; Molteni, Ryu \& Chakrabarti, 1996, hereafter, MRC96; 
Ryu, Molteni \& Chakrabarti, 1997, hereafter RMC97; Igumenshchev, Abramowicz \& Narayan 2000; 
Chakrabarti, Acharya \& Molteni, 2001, hereafter CAM01; Giri et al., 2010, hereafter GC10). 
However, these simulations were performed assuming that the flow has an equatorial 
symmetry, and therefore the flow behavior was studied only on one quadrant, i.e.,
first quadrant using the standard `reflection' boundary condition on the 
equatorial plane. In MLC94 and GC10, the results of standing and oscillating shock 
formations in inviscid flows are presented using Smoothed Particle Hydrodynamics (SPH) 
method and finite difference method respectively. In GC10, several simulations have been 
carried out choosing two conserved flow parameters (namely, specific energy ${\it E}$ 
and specific angular momentum $\lambda$) from the parameter space which provides complete 
set of solutions of a black hole accretion flow (C89). In order to break the reflection 
symmetry along equatorial plane, several simulations have been carried out by various groups 
for both black hole accretion (Molteni et al., 2001, hereafter M01;  Chakrabarti, Acharya \& Molteni, hereafter CAM01) as well as 
stellar wind accretion onto stars (Fryxell \& Taam, 1988 ; Taam \& Fryxell, 1989; Matsuda et al., 1991, 1992).
In M01 and CAM01, using SPH, it was shown that an instability
can occur in the flow. They also demonstrated that although matter is supplied symmetrically, those instabilities may not remain 
symmetric with respect to the equatorial plane. Furthermore, there is a strong interaction of the outgoing wind with the incoming flow 
(M01). However, SPH is known to be dissipative in nature and it is not impossible that in energy conserving schemes one might 
see that such oscillations are actually disrupting the flow altogether. We therefore extend the work of GC10 where energy 
is accurately preserved by removing the reflection condition along the equatorial plane. By this procedure, we intend to give 
answers to the following important questions: (a) Will the accretion flow be symmetric with 
respect to the equatorial plane? And if so, under what conditions? (b) Will this two quadrant flow have 
any effect on the formation of the so called `CENBOL'? This question is especially relevant 
as the CENBOL acts as the Compton cloud (Chakrabarti \& Titarchuk, 1995, hereafter CT95) 
while explaining the spectral and timing properties of black hole candidates.
(c) If the flow symmetry is absent then will the accretion flow remain stable 
at all or the flow would be violent and disrupted? (d) What will be its effects 
on outflows which are known to be produced on the CENBOL surface?

The plan of our paper is the following: in the next Section, we present the model 
equations governing the flow. In Section 3, we describe the methodology for our 
simulations using a grid based finite difference technique, called total variation diminishing 
(TVD) method which was discovered by Harten (1981). In Section 4, we discuss our 
simulation results and compare those with earlier simulations. Finally, 
in Section 5, we make concluding remarks.

\section{Model Equations without Viscosity}

In our approach, we consider a two dimensional axisymmetric flow around a Schwarzschild black hole. 
Instead of using general relativity, we use the pseudo-Newtonian potential first prescribed by 
Paczy\'nski and Wiita (1980). This potential mimics the effects of general relativity very successfully (see, RCM97, GC10). 
We use cylindrical polar coordinates ($r$, $\phi$ and $z$) for our calculations. 
The mass, momentum and energy conservation equations in a compact form using non-dimensional 
units are given in MRC96. We use the mass of the black hole
$M_{BH}$, the velocity of light $c$ and the Schwarzschild radius $r_g=2GM_{BH}/c^2$ 
as the units of the mass, velocity and distance respectively.   

The equations governing the inviscid flow have been presented in Ryu et al. (1995), MRC96 and GC10 in great detail 
and we do not repeat them here. In conservative form, the equations are given by,
$$
{\partial{\bmit q}\over\partial t}+{1\over r}{\partial\left(r
{\bmit F}_1\right)\over\partial r}+{\partial{\bmit F}_2\over\partial r}
+{\partial{\bmit G}\over\partial z} = {\bmit S},
\eqno(1a)$$
where the state vector is
$${\bmit q} = \left(\matrix{\rho\cr
                          \rho v_r\cr
                          \rho v_{\theta}\cr
                          \rho v_z\cr
                          E\cr}\right)_, $$
the flux functions are
$${\bmit F}_1 = \left(\matrix{\rho v_r\cr
                         \rho v_r^2\cr
                         \rho v_{\theta}v_r\cr
                         \rho v_z v_r\cr
                         (E+p)v_r\cr}\right)\qquad
{\bmit F}_2 = \left(\matrix{0\cr
                          p\cr
                          0\cr
                          0\cr
                          0\cr}\right)\qquad
{\bmit G} =  \left(\matrix{\rho v_z\cr
                         \rho v_r v_z\cr
                         \rho v_{\theta} v_z\cr
                         \rho v_z^2+p\cr
                         (E+p)v_z\cr}\right)_, \eqno(1b)$$
and the source function is
$$
{\textfont1 = \twelvei
      \scriptfont1 = \twelvei \scriptscriptfont1 = \teni
       \def\mit{\fam1}
{\bmit S} =  \left(\matrix{0\cr
                ~~~\cr
                {\rho v_{\theta}^2\over r}
                -{\rho r\over2\left(\sqrt{r^2+z^2}-1\right)^2\sqrt{r^2+z^2}}\cr
                ~~~\cr
                ~~~\cr
                -{\rho v_r v_{\theta}\over r}\cr
                ~~~\cr
                -{\rho z\over2\left(\sqrt{r^2+z^2}-1\right)^2\sqrt{r^2+z^2}}\cr
                ~~~\cr
                ~~~\cr
                -{\rho \left(rv_r+zv_z\right)\over
                2\left(\sqrt{r^2+z^2}-1\right)^2\sqrt{r^2+z^2}}\cr}\right)_.}
\eqno(1c)$$
Here, expression for energy density $E$ (without potential energy) is given by,
$$E=p/(\gamma-1)+\rho(v_r^2+v_{\theta}^2+v_z^2)/2,$$ $\rho$ is the mass density,
$\gamma$ is the adiabatic index, $p$ is the pressure, $\rho$ is mass density, 
$v_r$, $v_\theta$ and $v_z$ are the radial, azimuthal and vertical component of velocity           
respectively. In case of an axisymmetric inviscid flow, the equation for azimuthal 
component of the momentum simply signifies the conservation of specific angular momentum $\lambda$,
$$
\frac{d\lambda}{dt}=0.
$$

The general form of the equations of the flow in an inertial reference frame (Batchelor, 1967) is given by,
$$
\rho [{\partial {\bf v} \over \partial t} + {{\bf v} . {\nabla {\bf v}}}] = 
- { \nabla p} + {\bf {F_b}}  + {\nabla . {\bf {\tau}}}, \eqno(2)
$$
where, ${\bf v}$ is the flow velocity, ${\bf \tau}$ is the viscous stress tensor, and ${\bf {F_b}}$  
represents body forces (per unit volume) acting on the fluid and ${\nabla}$ is the Del operator.
Typically, the body forces consist of only gravity forces, but may include other
types (such as electromagnetic forces). Since, here we are considering only inviscid and non-magnetic flow, 
detailed discussion on viscous stress tensor is out of the scope of this paper.

To describe the gravitational field around a Schwarzschild Black hole, we use the Pseudo-Newtonian gravitational 
field of a point mass $M_{BH}$ located at the centre in cylindrical coordinates $[r,\theta,z]$ described 
by Paczy\'{n}ski \& Wiita (1980) potential,
$$
\phi(r,z) = -{GM_{BH}\over2(R-r_g)}, \eqno{(4)}
$$
where, $R=\sqrt{r^2+z^2}$. 

\section{Methodology and simulation setup}
\begin{figure}
\begin{center}
\includegraphics[width=8cm]{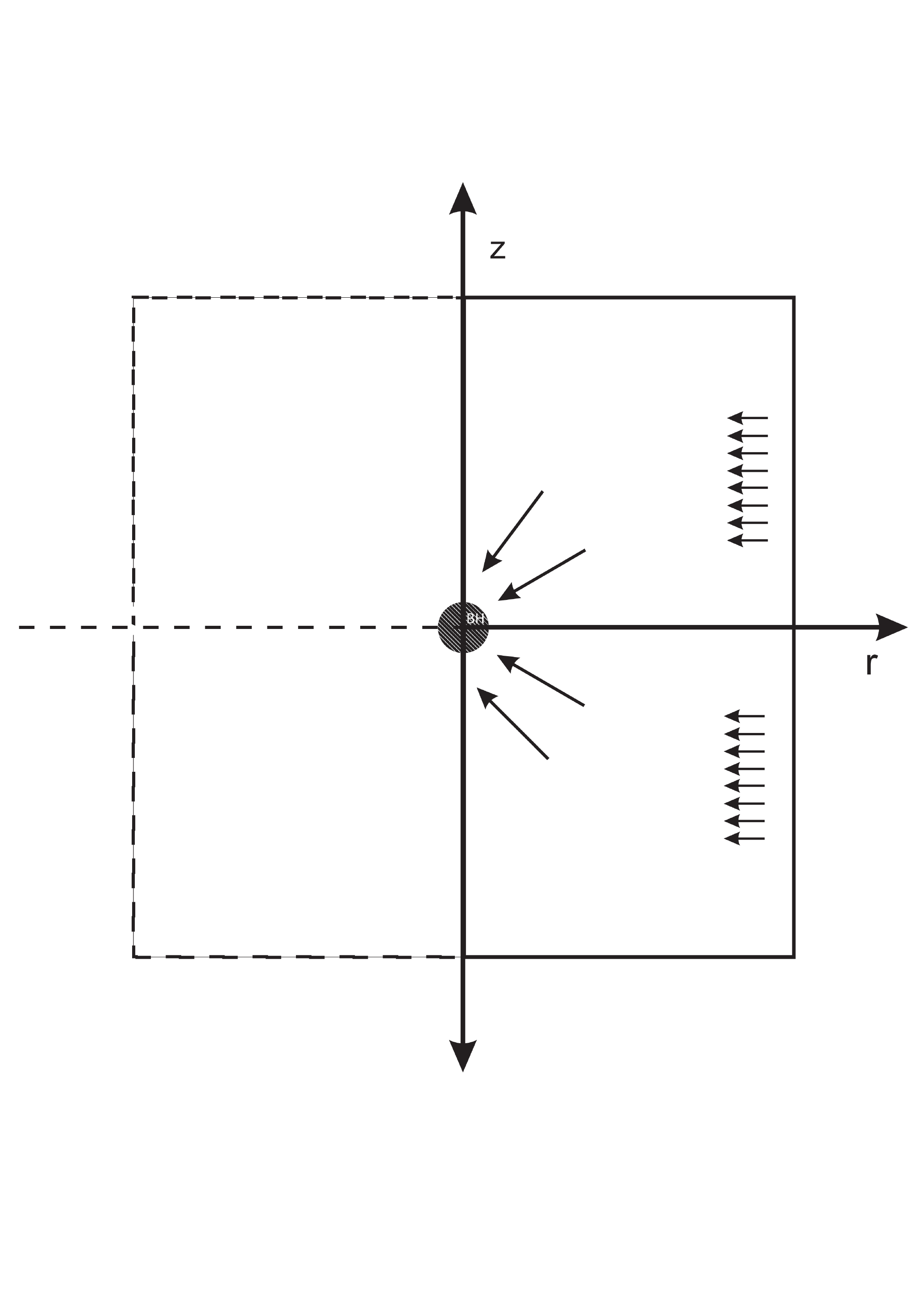}
\caption{A schematic diagram of the system under consideration. Solid box is our computational 
region having a dimension of $0 \leq r \leq 200$ and $-200 \leq z \leq 200$ on the $r-z$ plane.
Black hole is sitting at the origin of $r$ \& $z$ axis. Matter is injected from 
the outer boundary and matter is sucked into the black hole. No reflection symmetry along 
the equatorial plane has been assumed.}
\end{center}
\end{figure}
The computational box occupies two quadrants (first and fourth quadrant) of the r-z plane with $0 \leq r \leq 200$ 
and $-200 \leq z \leq 200$. Here, we use the reflection boundary condition only on z-axis to 
obtain the solution in other quadrants. This is because in all our simulations we assume 
the flow to be axisymmetric in nature as the black hole itself is strictly 
axisymmetric and the inner boundary condition is expected to force the flow to be axisymmetric close to the black hole. 
The incoming gas enters the computational box through the outer boundary (having a vertical cylinder like shape
in three-dimensions), located at $r_b = 200$. The flow is injected symmetrically both above and below the equatorial plane. 
We supply radial velocity $v_r$, the sound speed $a$ (i.e., temperature) of the flow at boundary points 
from the hybrid model and boundary values of density $\rho$ from standard vertical equilibrium solution (C89). 
We scale the density in such a way that the incoming gas has the density of ${\rho}_{in} = 1.0$. 
In order to mimic the horizon of the black hole, we place an absorbing inner boundary at $R = 2.5 r_g$, 
inside which all the incoming matter is absorbed completely into the black hole. The inner sonic point 
is formed around this radius anyway, so this choice of inner boundary does not affect the flow dynamics 
upstream. In order to avoid singularities caused by `division by zero', we fill the grid with a 
background matter of very low density ${\rho}_{bg} = 10^{-6}$ having a sound speed (or, temperature) 
to be the same as that of the incoming matter. Hence, the incoming matter has a pressure $10^6$
times larger than that of the background matter. Of course, this initial matter is totally washed out
and replaced by the injected matter within a dynamical time scale. 
Initially, the low density matter with which the grids are filled, is assumed to be static, i.e., 
the values of radial ($v_r$), rotational ($v_\phi$)  
and azimuthal ($v_z$) components are all chosen to be zero for all 
the grids except those on the outer boundary. Thus, the Mach number 
is zero everywhere except on the outer boundary at the beginning of the simulation.
The calculations were performed with a very high resolution $512 \times 1023$ grids. Thus,
each r and z-grid has a size of $0.3906$ in units of the Schwarzschild radius. 
Figure 1 shows schematically our system on the $r-z$ plane in a cylindrical co-ordinate system.
In this work, we are interested to see time dependence of CENBOL and instability in the 
accretion disk around a black hole. So, the resolution which we have ($ \sim 0.4$ Schwarzschild radii)
is enough to catch these salient features. 
All the simulation cases have been carried out assuming a stellar mass black hole $M_{BH}= 10{M_\odot}$. 
The conversion of our time unit to physical unit is $2GM_{BH}/c^3$, and thus 
the physical time for which the programme was run would scale with the mass of the black hole.
We typically find that the infall time from the outer to the inner boundary is about $\sim 0.5$s. This is  
computed from the sum of $dr/<v_r>$ over the entire radial grid, $<v_r>$ being averaged over $20$ vertical grids. 
We carry out simulations for several hundreds of dynamical time-scales in order to stay away from any transient effects.

\section{Results} 
\begin{figure}
\begin{center}
\includegraphics[width=8cm]{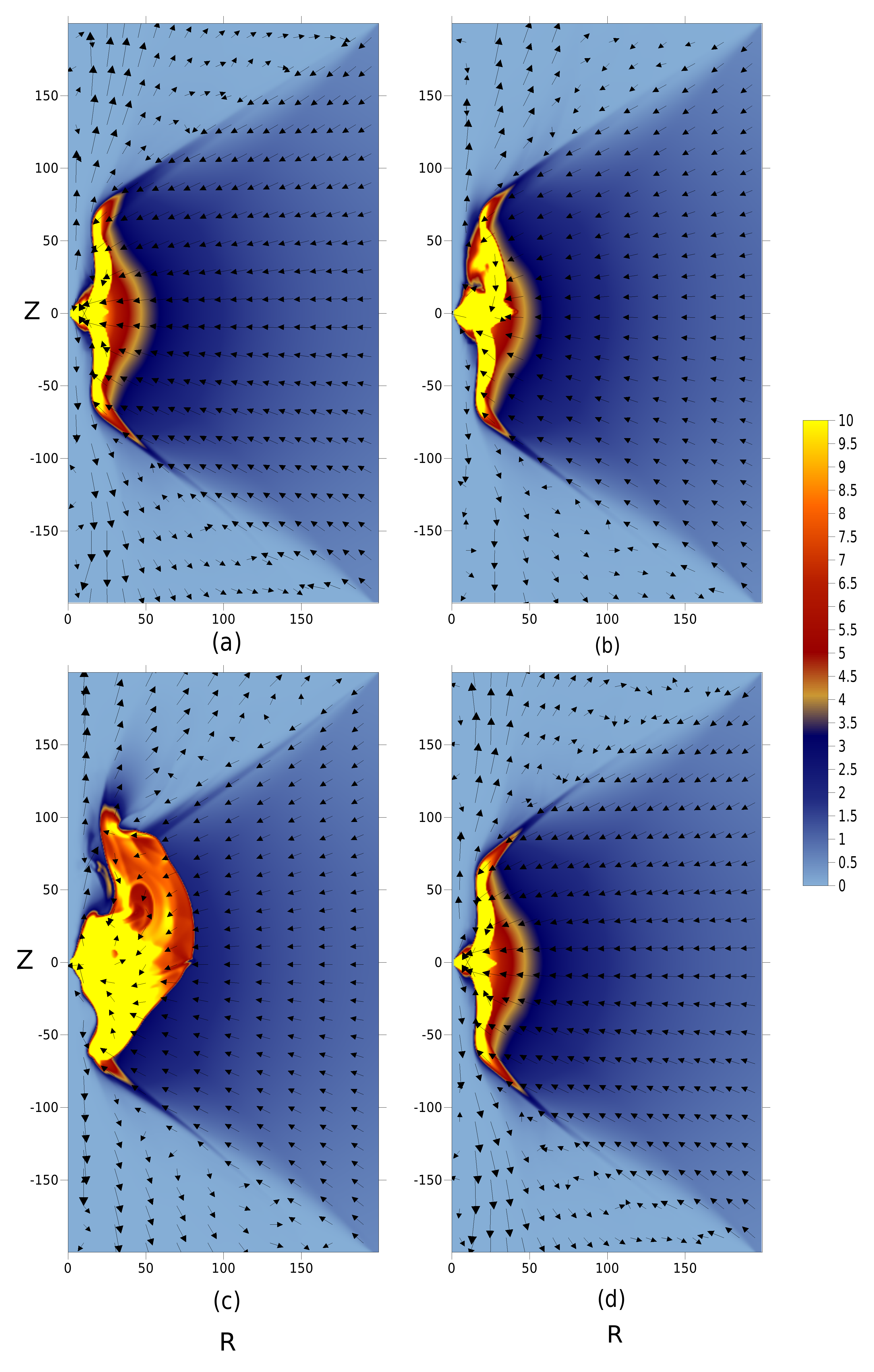}
\caption{Density and velocity vector map to show quasi-periodic formation and 
deformation of the CENBOL at (a) t = 21.36 s, (b) 21.84 s, (c) 22.08 s and (d) 24.95 s. Specific angular 
momentum is considered to be $\lambda = 1.6$. In (a), high density region due to centrifugal supported boundary
layer (CENBOL) has a symmetric shape. In (b), symmetry is about to be broken due to vertical oscillation of the 
perturbing mass. In (c), CENBOL is deformed but not destroyed. In (d), the CENBOL is restored back. }
\end{center}
\end{figure}
\begin{figure}
\hfill
\subfigure[]{\includegraphics[width=8.5cm]{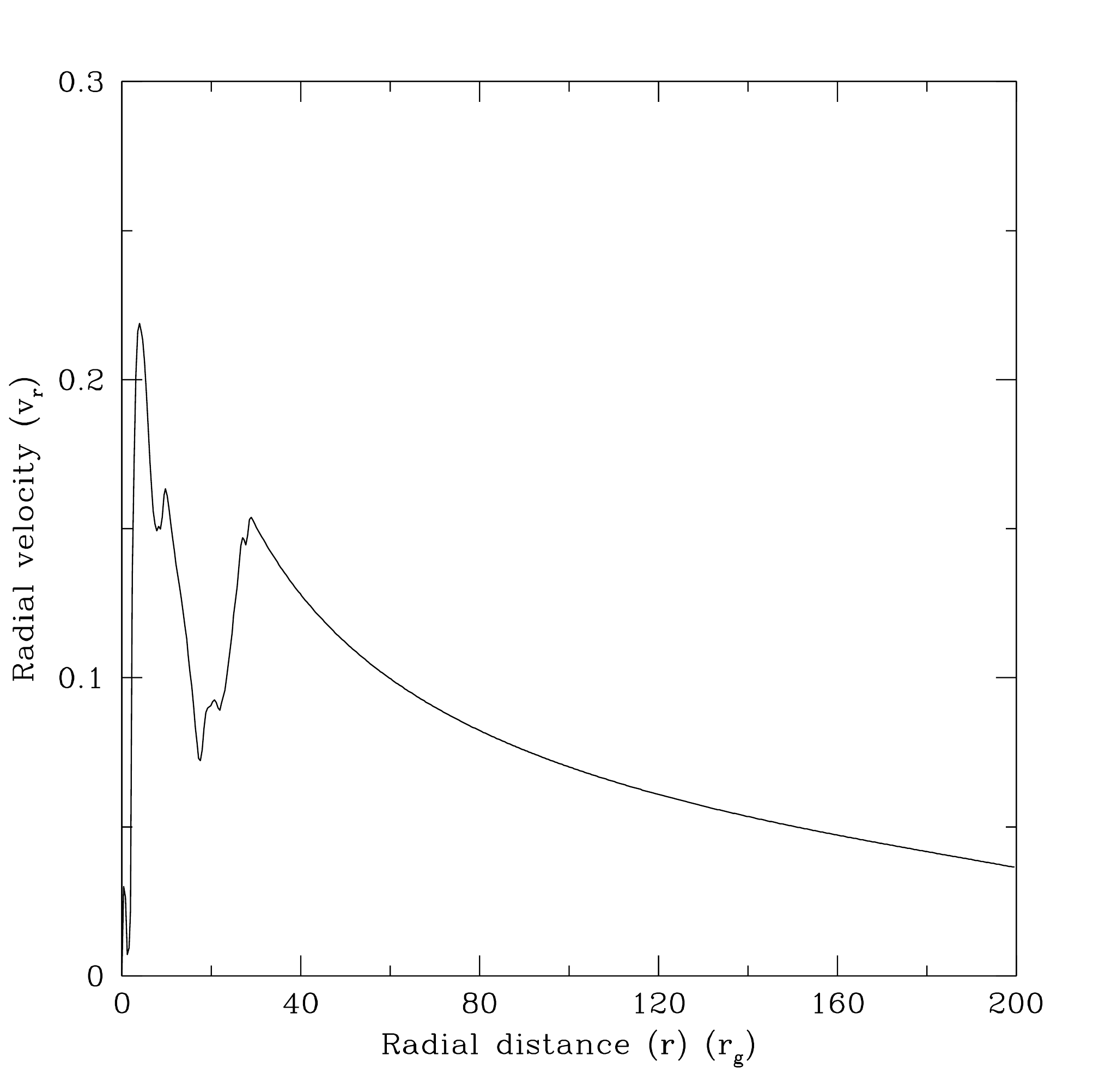}}
\hfill
\subfigure[]{\includegraphics[width=8.5cm]{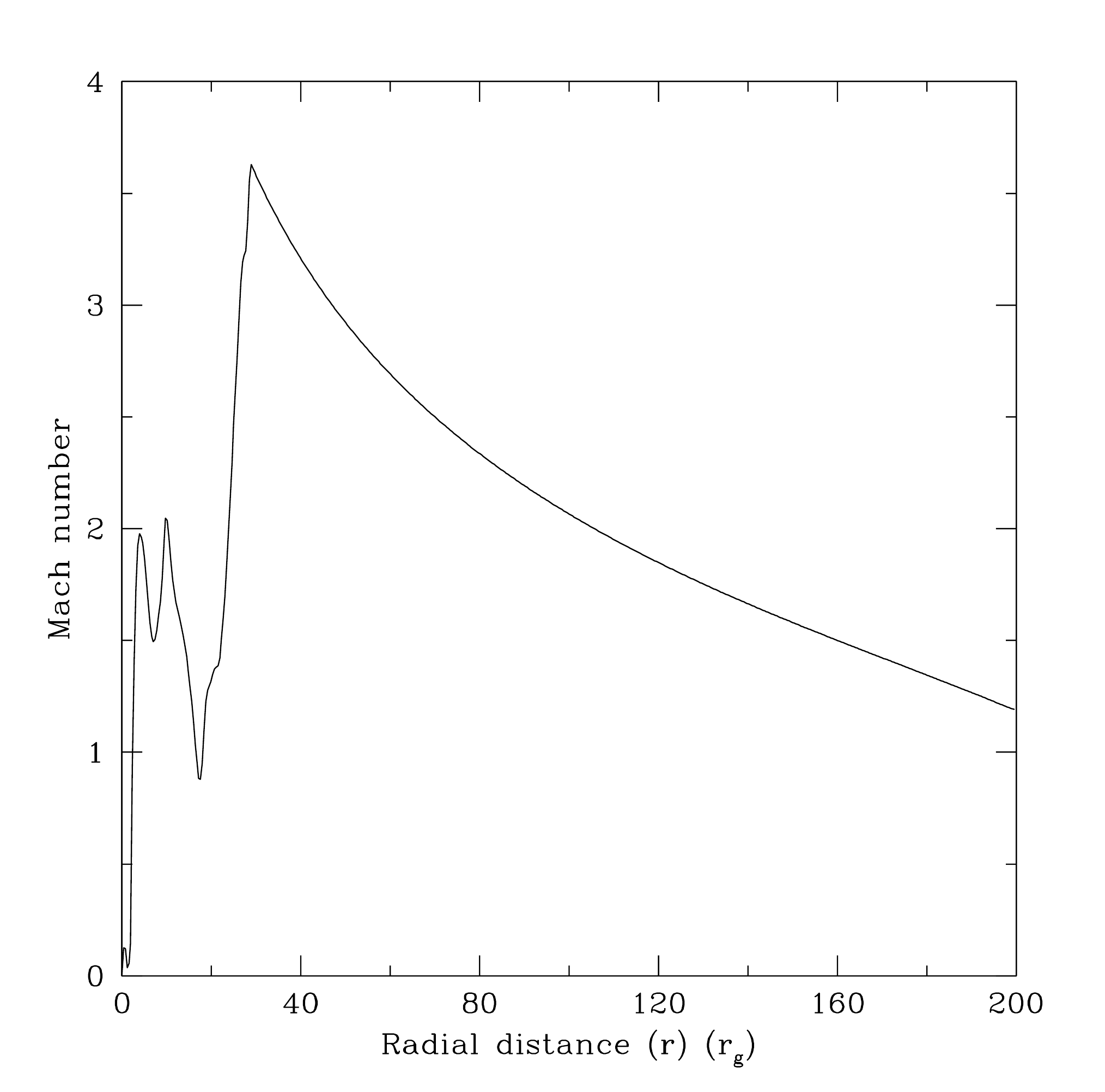}}
\hfill
\caption{Radial distribution of the (a) radial velocity component and 
(b) radial Mach number ($v_r/a$) on the equatorial plane. Time is $t = 24.95$ s and specific 
angular momentum is $\lambda = 1.6$. We clearly see the slowing down of matter at the centrifugal 
barrier (a) and a supersonic to sub-sonic transition (b) forming a shock.}
\end{figure}   

GC10 carried out all their simulations assuming a reflection boundary condition on the equatorial plane and injected matter 
only in one quadrant. M01, in their work with cooling processes, has shown that certain instabilities in the 
accretion disk set in even in SPH simulations where angular momentum is preserved more accurately (see, MRC96, RCM97). 
Chakrabarti et al. (2004) have shown the quasi-periodic variabilities caused due to vertical 
and horizontal oscillation of shock waves in a two dimensional axisymmetric flow using Smooth 
Particle Hydrodynamics (SPH). Our result in this paper, obtained using a code
with finite difference method which employs an accurate Total Variation Diminishing (TVD) scheme, 
clearly supports this work. Following C89, we injected the flow at the outer boundary with matter 
in vertical equilibrium. The injection rate of the momentum density is kept uniform throughout the injected height 
at the outer edge. We stop the simulations at $t=95$s (physical time). This time is more than two 
hundred times the dynamical time of the flow. Solutions presented in Fig. 2(a-d) are at 45-50 
times the dynamical time, long after transient effects ($\sim 1$s) die out. Thus the effects seen are real 
and are expected to influence spectral and timing properties significantly.\\
C89 predicted that the standing shocks can form if $\lambda >1.525$. We find that, indeed, CENBOL is produced 
when $\lambda>1.5$, the `discrepancy' being perhaps due to the presence of turbulence pressure (generated by 
the centrifugally bounced outward flow colliding with the infalling gas in post-shock region) 
which helps the formation of CENBOL even at a lower angular momentum. 
\begin{figure}
\begin{center}
\includegraphics[width=9.0cm]{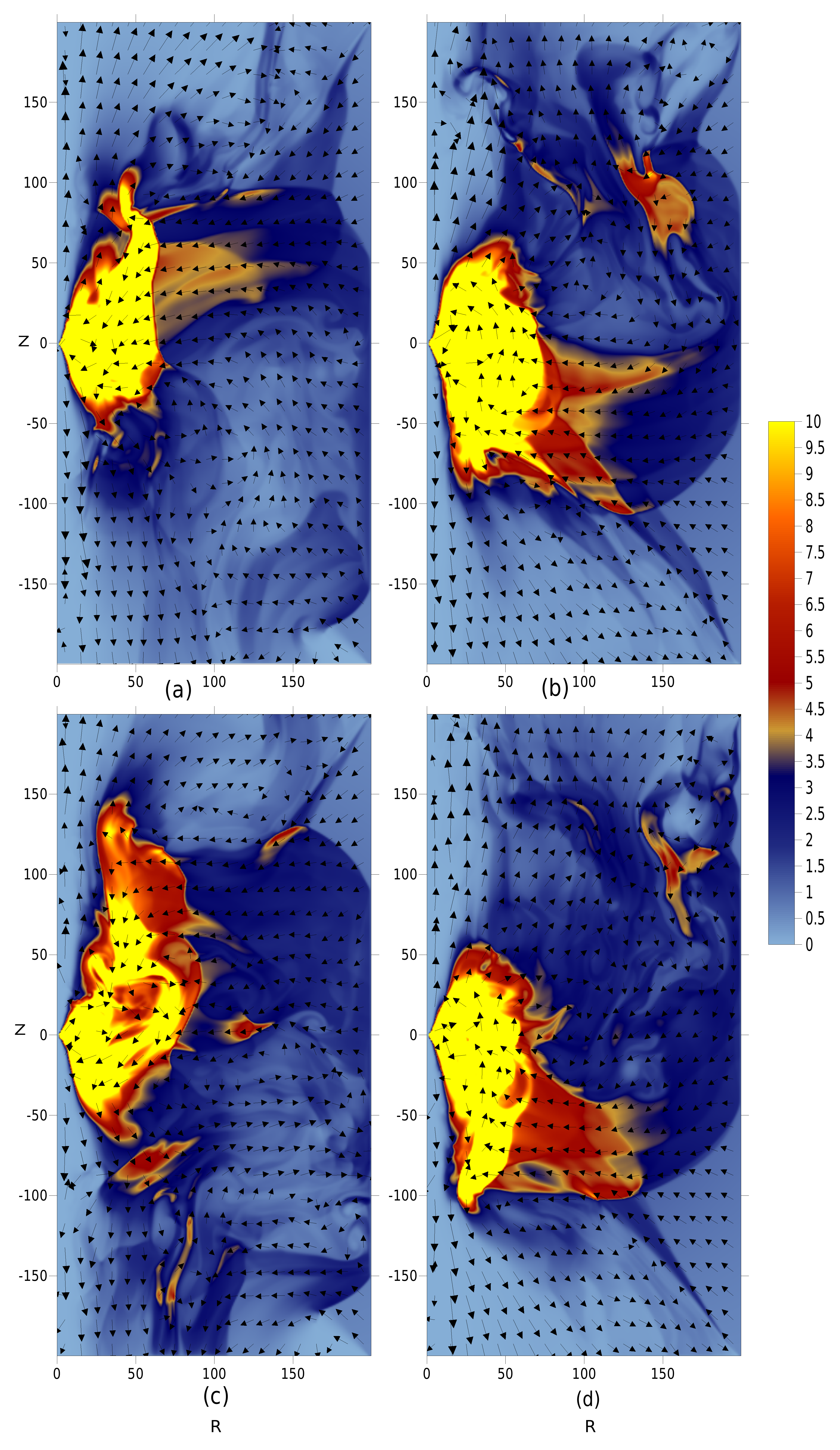}
\caption{Density and velocity vector plots of the accretion flow to show instability 
within the flow. (a) and (c) show that the CENBOL is shifted above: flow density is higher in the upper quadrant 
and (b) and (d) show that the CENBOL is shifted below: flow density is higher in the lower quadrant. 
Plots are drawn at $t= 17.34, 22.57, 40.08$, and $43.2$ seconds respectively. 
Specific angular momentum is considered to be $\lambda=1.7$.}
\end{center}
\end{figure}
In Fig. 3, we show the distribution of radial component of the velocity and the Mach number which is the ratio 
between radial velocity ($<v_r>$) and sound speed ($<a>$). Velocity and sound speed are averaged over $10$ grids 
located on both sides of the equatorial plane. The plot is drawn for $t \sim 24.95$ s. The radial velocity 
suddenly drops at the shock location where the flow also becomes subsonic as evidenced by the Mach number distribution.
As the time progresses, the shock tends to oscillate back and forth and also vertically.\\
\begin{figure}
\subfigure[]{\includegraphics[width=7.0cm]{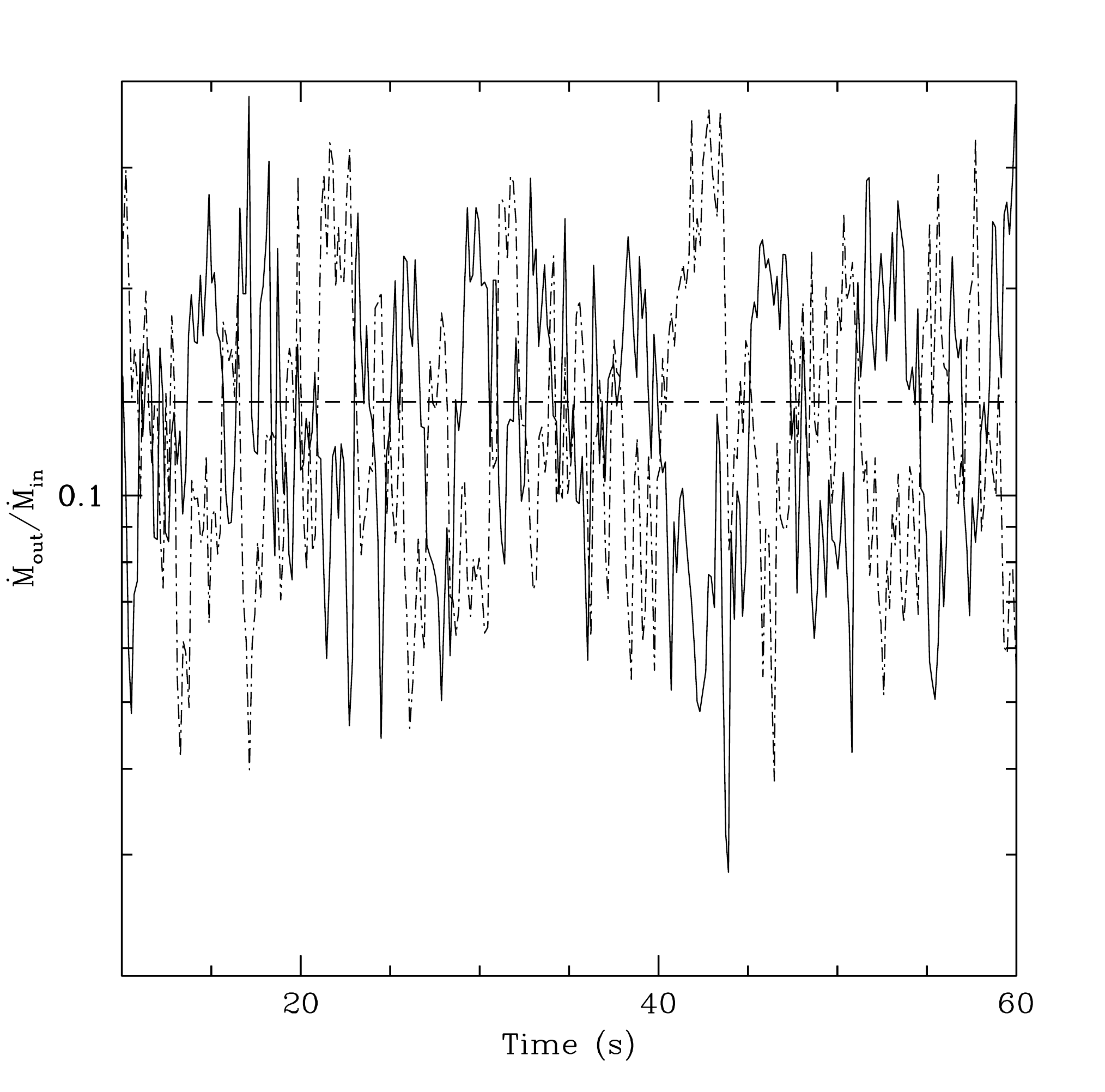}}
\subfigure[]{\includegraphics[width=7.0cm]{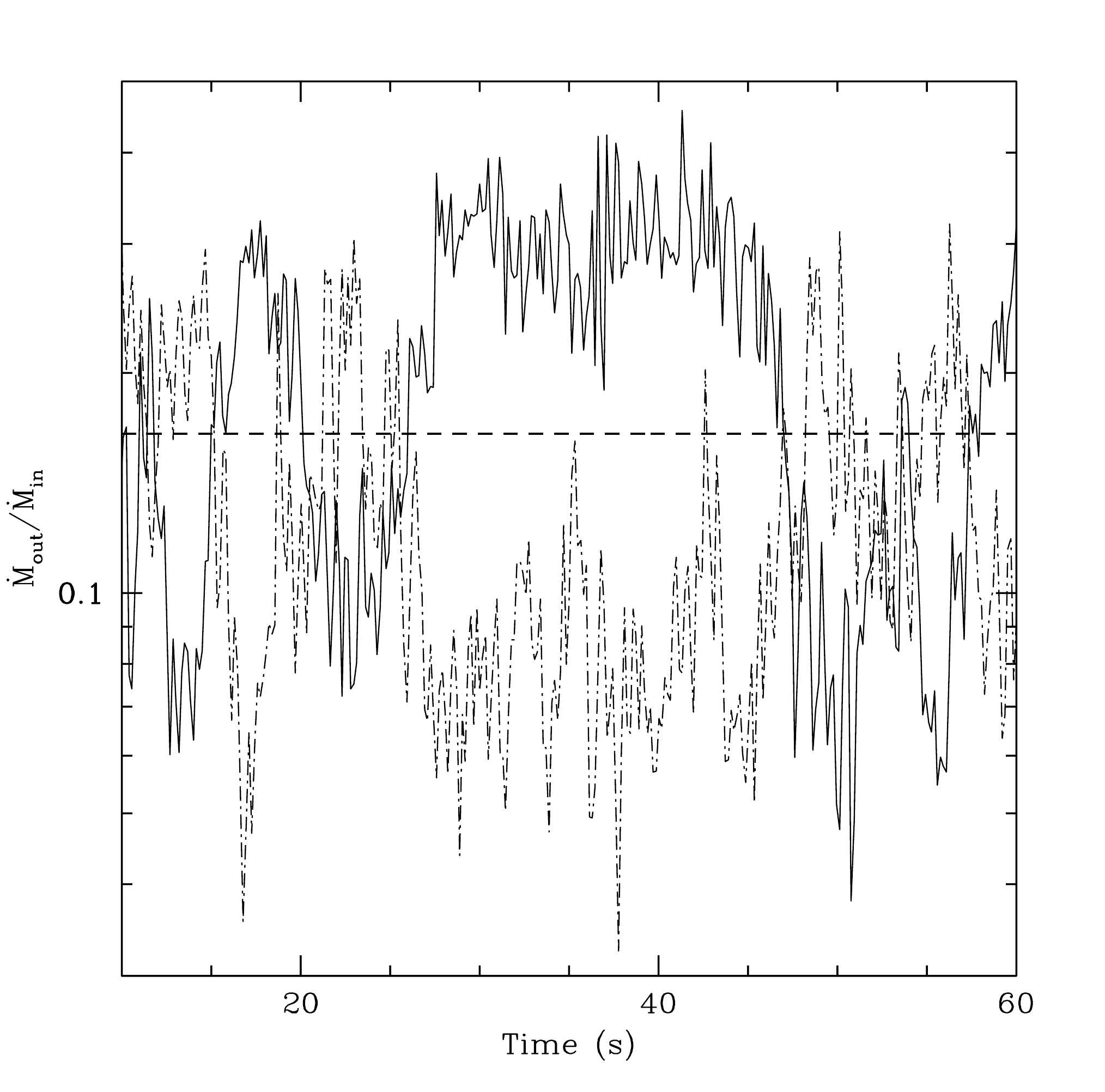}}
\caption{Time variation of the ratio between the total outflow rate
(${\dot M}_{out}$) and the total inflow rate (${\dot M}_{in}$) 
showing an anti-correlated behaviour. When the outflow rate from the upper boundary is high,
the outflow rate in the lower boundary is low and vice versa. 
Here (a) $\lambda = 1.7$ and (b) $\lambda = 1.8$. Dot-dashed curve represents the outflow inflow rate ratio in upper quadrant, 
solid curve represents the ratio in lower quadrant, and dashed curve represents the mean value (time averaged) 
of the outflow inflow rate ratio. We note that the mean outflow rate as well as the degree of deviation from the mean is higher when 
angular momentum is higher.}
\end{figure} 
We now carry out a simulation with a larger specific angular momentum. We choose $\lambda = 1.7$. 
The centrifugal force increases the location of the shock and thus the size of the
CENBOL is increased. However we notice that after a transient state, the CENBOL takes part in a vertical oscillation around the 
equatorial plane. We also observe that the outflowing wind is interacting with the incoming accreting matter creating 
weaker oblique shocks near top-right and 
bottom-right corners. In Fig. 4(a-d), we plot density and velocity vector maps of the flow at $t = 17.34$ s, $22.57$ s, 
$39.13$ s, and $43.23$ s respectively. The CENBOL, though distinct, takes a complex shape.
It is also evident that the outflow oscillates, i.e., there are times when outflow from the upper 
quadrant is large and at other times the outflow in the lower quadrant is large. 
Disks in Fig. 4(a) and Fig. 4(c) have the CENBOL 
shifted slightly to upper quadrant and the outflow rate from the upper boundary is found to be larger. However, 
disks in Fig. 4(b) and Fig. 4(d) have the CENBOL shifted slightly towards the lower quadrant and the outflow from the lower 
boundary is found to be larger. 
\begin{figure}
\subfigure[(i)]{\includegraphics[width=8.3cm]{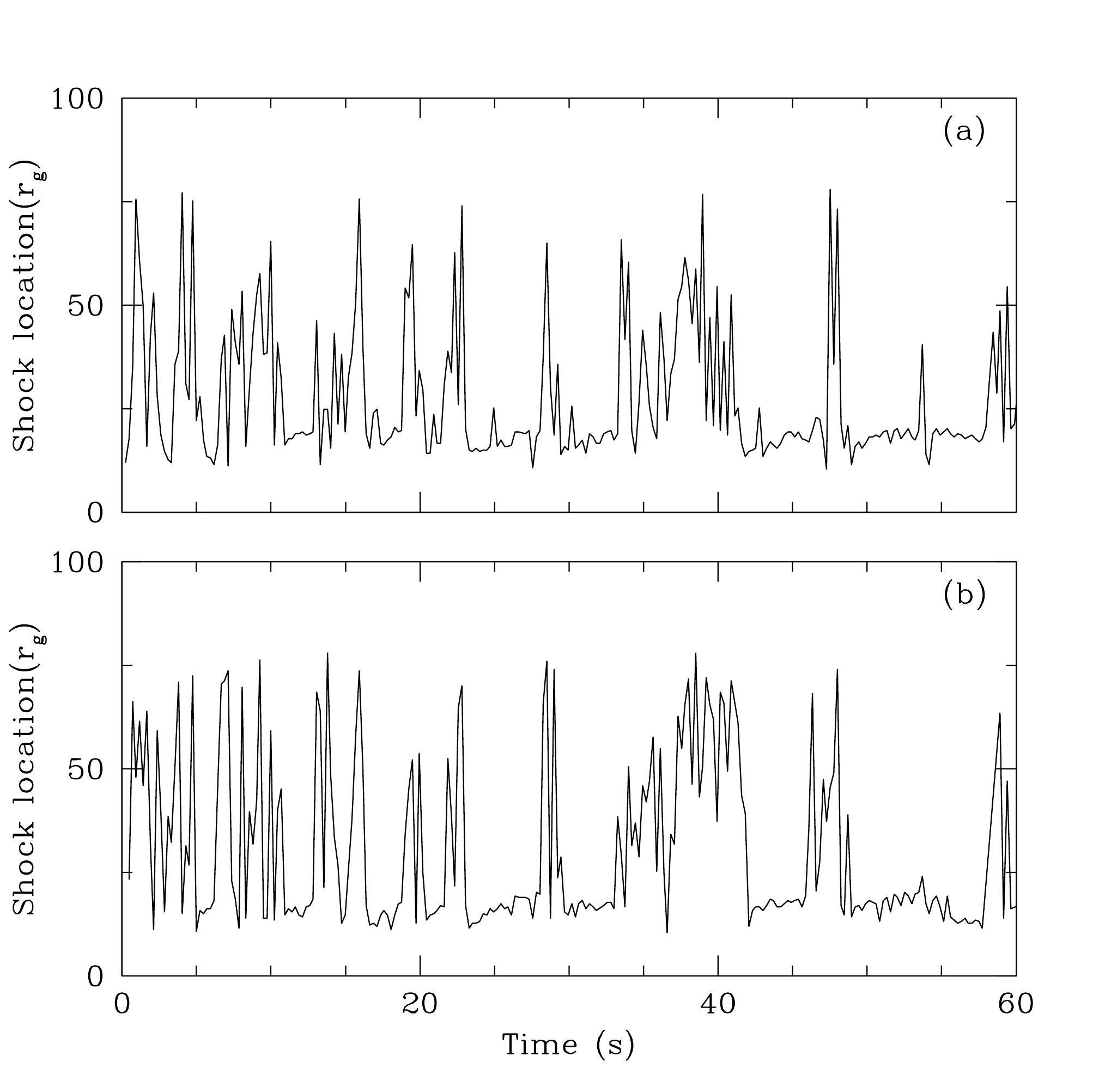}}
\subfigure[(ii)]{\includegraphics[width=8.3cm]{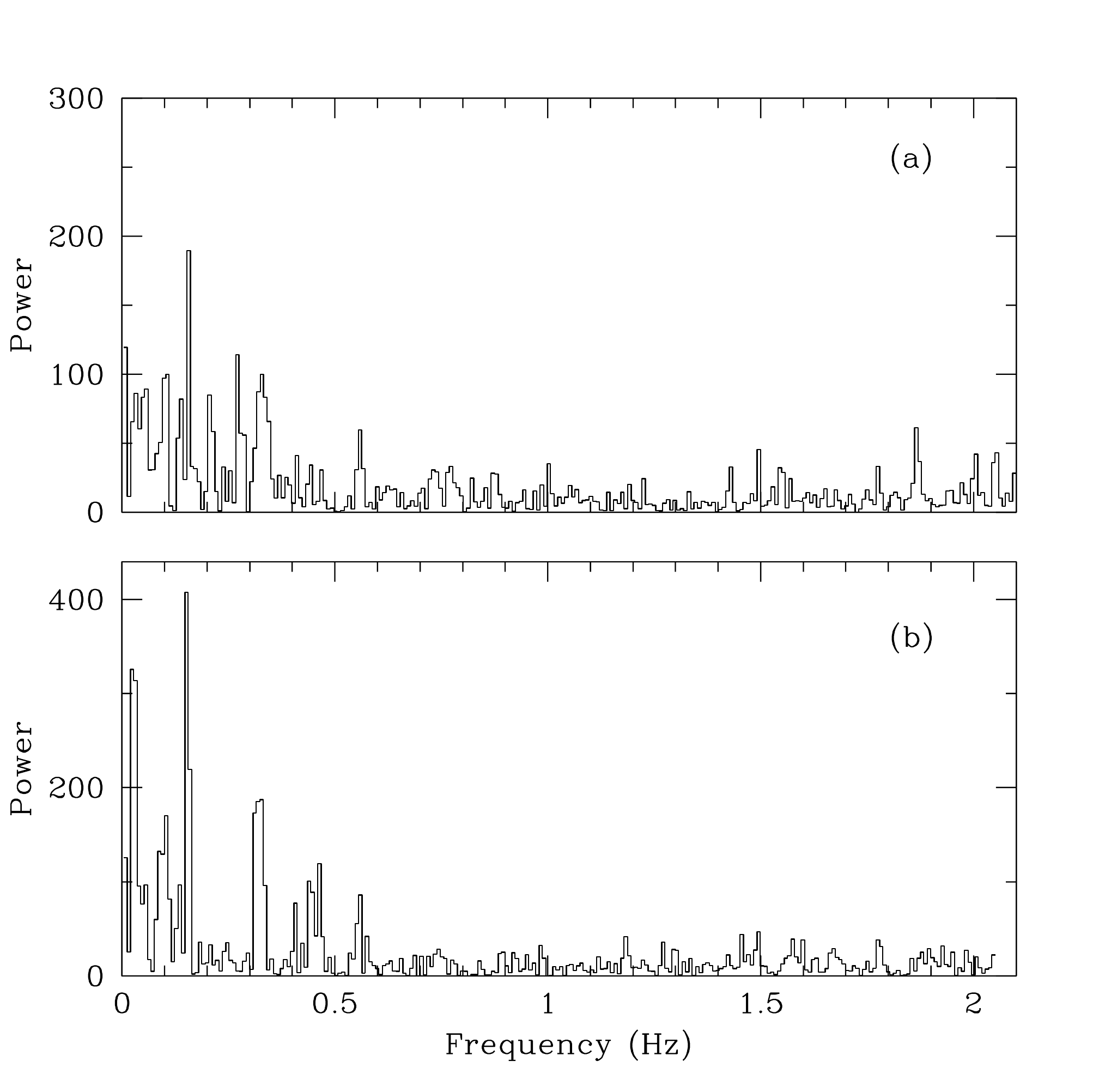}}
\caption{Time variation of shock location for two quadrant flow with $\lambda = 1.6$
angular momentum and their respective power density spectra.  Group (i) show the shock location
variation in upper (a) and lower (b) 
quadrants. Group (ii) show the power density spectra (PDS) of these locations. We see evidence of a prominent peak at 
$\sim 0.16$Hz in both the cases and a harmonics at $\sim 0.32$Hz. 
}
\end{figure} 
\begin{figure}
\subfigure[(i)]{\includegraphics[width=8.3cm]{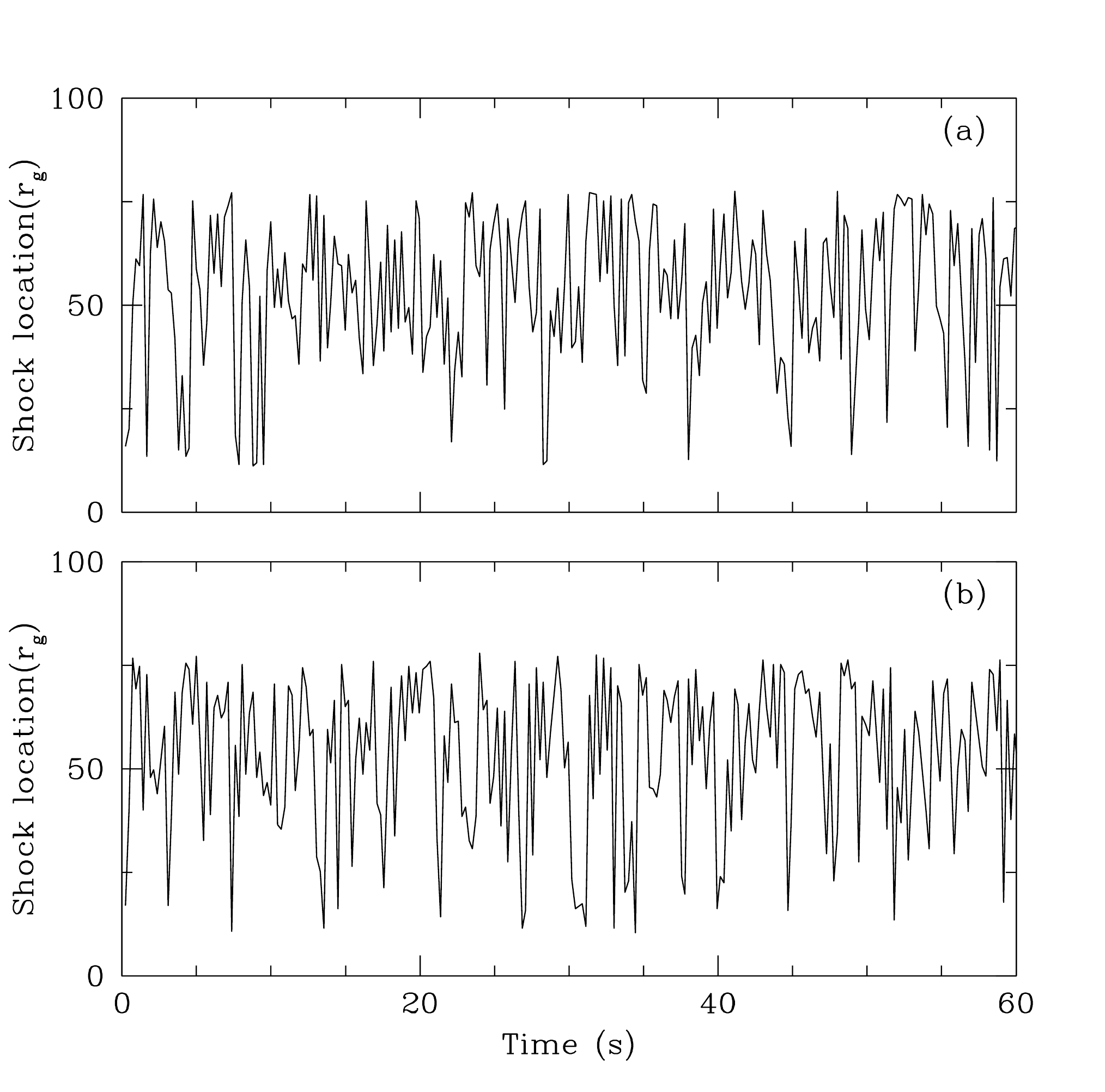}}
\subfigure[(ii)]{\includegraphics[width=8.3cm]{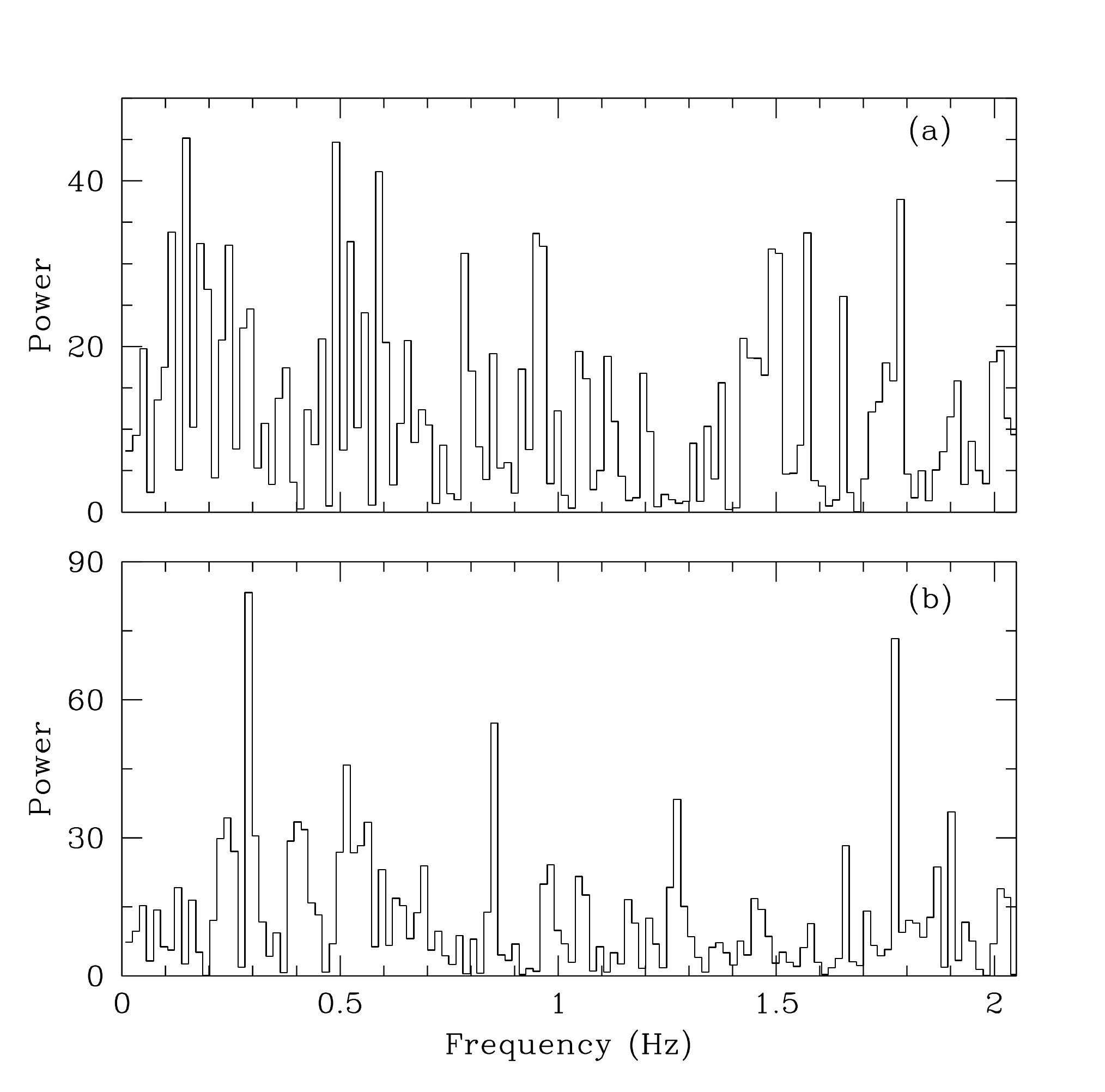}}
\caption{Same as in Fig. 6 for  $\lambda = 1.7$. Note that the shock  oscillates around a larger mean location, though the
symmetry in upper and lower quadrant is lost.  The power density spectra have several peaks and the oscillation is more chaotic. }
\end{figure}
\begin{figure}
\subfigure[(i)]{\includegraphics[width=8.3cm]{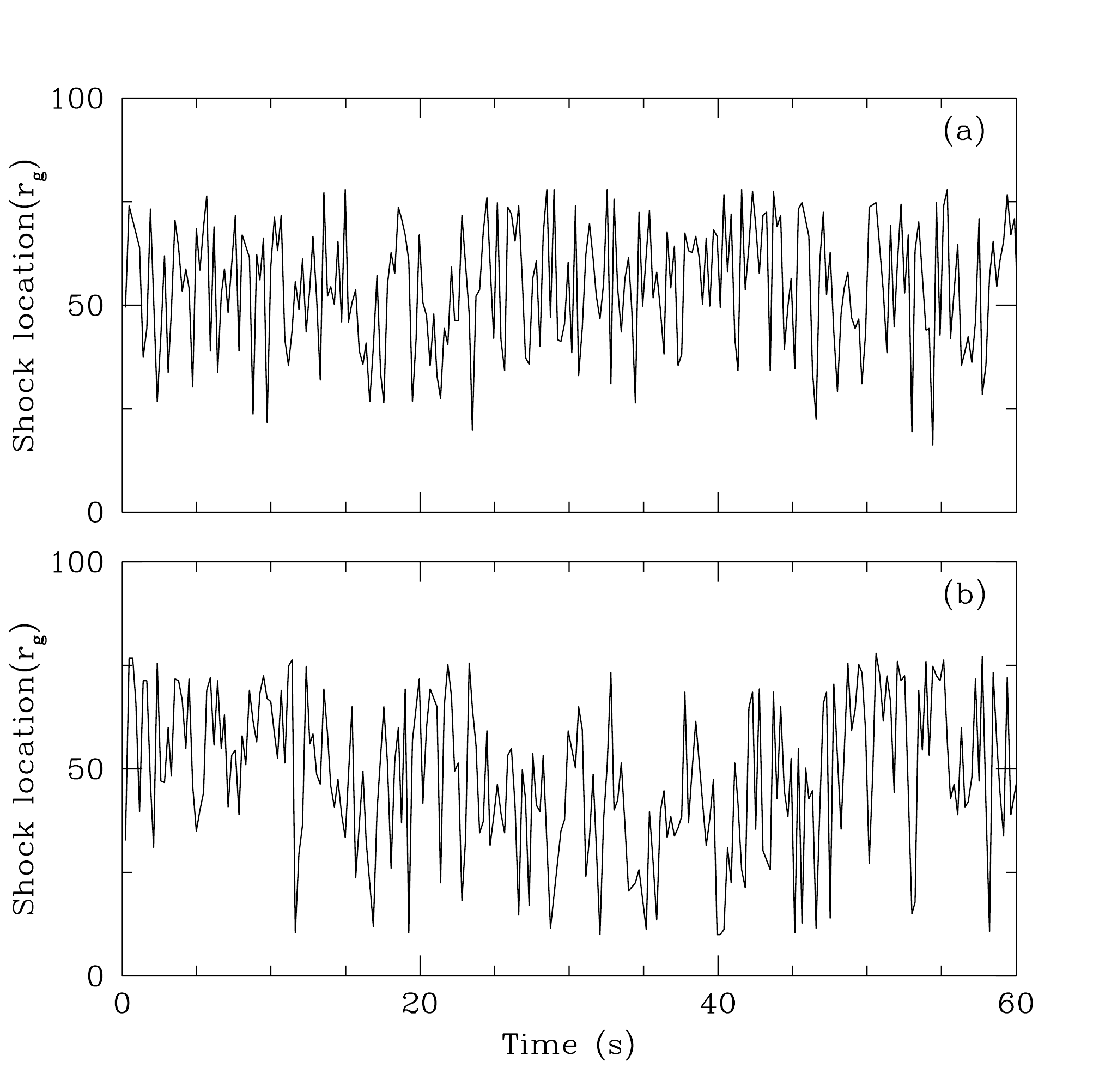}}
\subfigure[(ii)]{\includegraphics[width=8.3cm]{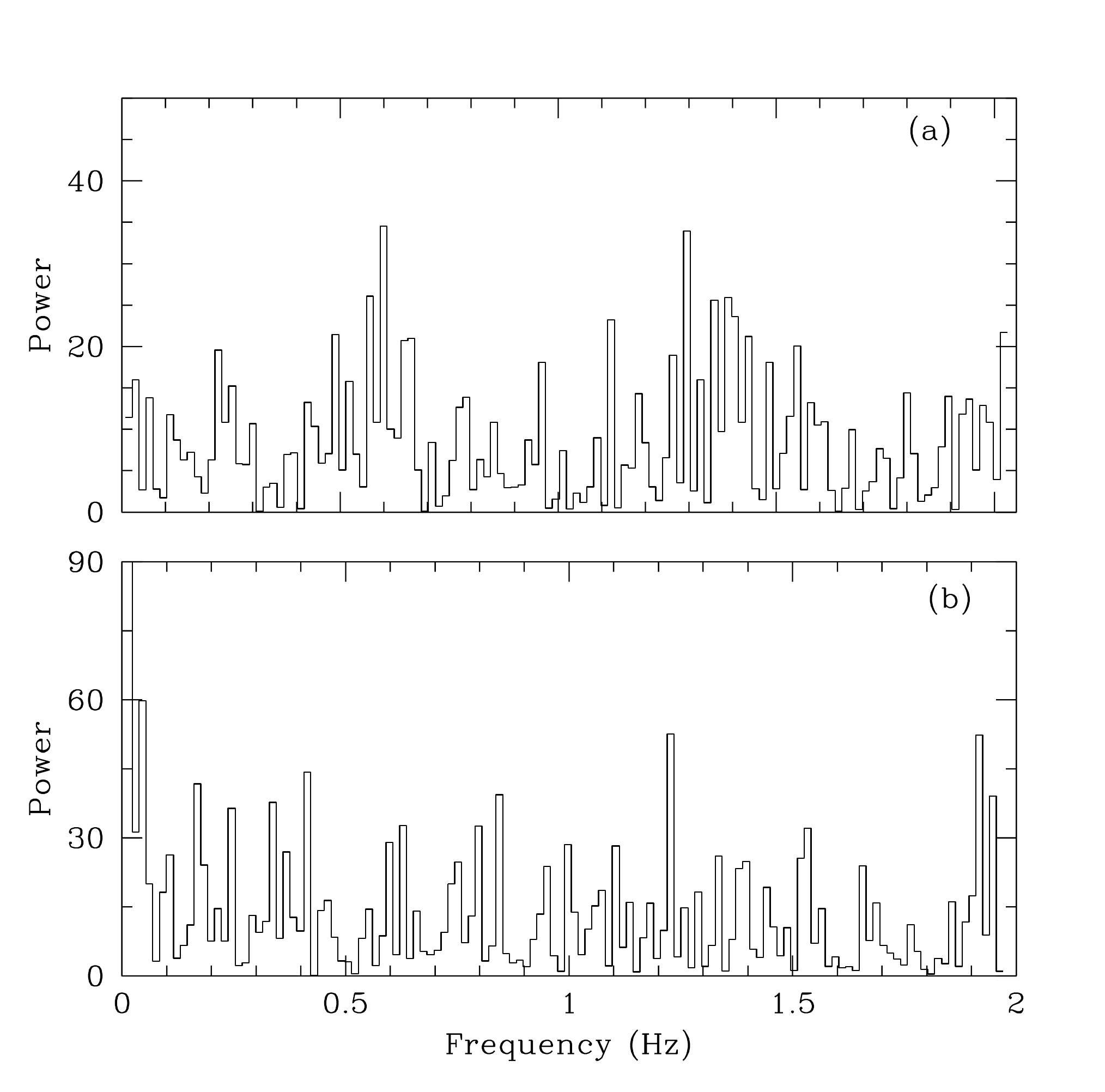}}
\caption{Same as in Fig. 6 for  $\lambda = 1.8$. Note that the shock oscillates around a mean location similar to what we observed 
for $\lambda=1.7$, perhaps due to the post-shock turbulences. The symmetry 
in upper and lower quadrant is lost. The power density spectra have several peaks and the oscillation is more chaotic. }
\end{figure} 
\\In Fig. 5, we plot the ratio of the outflow rate to constant injected rate. The dot-dashed curve represents 
the ratio of the total outflow rate (${\dot M}_{out}$) and the total inflow rate (${\dot M}_{in}$) in the 
upper quadrant, solid curve represents the ratio in lower quadrant, and dashed curve represents the mean value (time averaged)
of the ratio. In Fig. 5(a) and Fig. 5(b), we
use $\lambda=1.7$ and $1.8$ respectively. We note that outflow rate as well as the mean rate increases with
specific angular momentum. This further establishes that the outflow is centrifugally driven. The overall
rate is found to be about $5-10$\% of the inflow rate. We also note that the deviation from the mean rises
as well, indicating that for higher $\lambda$, stronger vertical oscillation sets in. This was also seen in the density and velocity plots.
These plots show highly anti-correlated behaviour between the rates from the upper and the lower boundaries. 

\section{Discussions and concluding remarks} 
In this paper, we  presented extensive time dependent numerical simulations
of two quadrant accretion flow around a black hole.
In earlier studies, such as in MLC94 and GC10, similar type of simulations were carried out in one quadrant using SPH and TVD schemes respectively. In these earlier simulations, our goal was to check if the shocks could be produced in the first place, and if yes, how does the puffed up post-shock region behave in reference of thick accretion flow. 
It was found that indeed, since the post-shock flow is sub-sonic with low radial velocity and sub-Keplerian angular momentum, it does behave as a thick accretion disk, though without a cusp since radial velocity increases close to the inner edge. This post-shock region (CENBOL) was then used to Comptonize the low energy photons from a Keplerian disk. 
In the present work, we asked ourselves if the CENBOL really remains symmetric with respect to the equatorial plane. For this we removed the reflection symmetry imposed forcefully in earlier simulations. We inject matter only in first and the
fourth quadrants. We find that for lower centrifugal force, the CENBOL remains symmetric, though a vertical oscillation sets it which becomes more and more violent as the specific angular momentum increases. This is superimposed with a horizontal oscillation. We also find that the outflow rate from each of the quadrants independently vary: The quadrant in which the CENBOL is tilted, also has the higher rate. Thus, the rates in the two quadrants are anti-correlated.

 There are two important aspects which need major discussions: First, the oscillations seen here are in radial 
and vertical directions. In earlier works, especially in Ryu et al. (1997), it was shown that when the Rankine-Hugoniot
relations are not satisfied, the shocks undergo radial oscillations. This is because a transonic accretion flow with a 
significant amount of angular momentum, has two physical sonic points and the flow is required to have higher 
entropy in order to pass through the innermost one (Chakrabarti, 1989). This means that even when the conditions 
of steady shock formation is not satisfied, the flow will pass through the inner sonic point and the shock will be unsteady, moving radially, and searching for an acceptable solution. Another reason of radial oscillation was found by Molteni et al. (1996a) where it was seen that shocks start to oscillate only when the cooling time scale in the post-shock region roughly matches with the 
infall time scale in that region. Since cooling process is absent in the present simulation, this latter explanation is
not relevant in the present context. In presence of viscosity, oscillation due to viscous overstability (Kato, 1978) is
well known. But ours is an inviscid flow. Hence this is also not the reason of the oscillation seen in our simulations. In case of slender tori, inertial modes are rapidly excited faster than the dynamical time scale (Blaes, 2006) and causes significant instability
(Horak, 2012). However, our post-shock region is advecting and not slender. Perturbations due to the inertial modes 
are likely to be advected out of the disk when the flow passes through the innermost sonic point. 

In the present context, we note that the disk instability is high only when the angular momentum is very large. 
Two important physical processes are triggered by angular momentum: (i) infalling matter hits the centrifugal barrier
(defined by the location where the centrifugal forces is similar to the gravitational force) and bounces back near the 
equatorial plane. This flow confronts the incoming matter and two turbulence cells of opposite vorticity are generated, one above and the other below the equatorial plane. (ii) Centrifugal
pressure driven winds are formed which also flow outwards (in between the centrifugal barrier and 
the so-called funnel wall, see, Molteni et al. 1996b), confronting the incoming flow away from the 
equatorial plane. This interface is therefore susceptible to Kelvin-Helmholtz instability. In all the cases we ran, 
we found that for very low angular momentum, the wind does not form at all and thus this instability is absent. Higher the angular momentum, stronger is the shear instability between the incoming and outgoing components. When the amplitude of the fastest growing mode becomes non-linear, instabilities in the upper and lower halves join and push the entire disk on one side or the other. This is what we believe to be the cause of the vertical motion. Clearly, this requires a thorough study.

The simulations we carried out are inviscid and thus the oscillations are more violent as the angular momentum is not transported away. Similarly, we have not included radiative cooling, because of which, the flow is very hot inside CENBOL. It has already been suggested that the shock oscillation contributes to quasi-periodic oscillations (QPOs) of black hole candidates. We believe that the vertical oscillation of CENBOL may contribute to the variability classes as seen in objects such as GRS 1915+105. The vertical erratic movement of the CENBOL could be responsible for the varying Comptonized component as seen by a specific observer and this may give rise to a large number of variability classes even in so-called hard states, soft states or intermediate states. 

Giri \& Chakrabarti (2013) and Giri et al. (2015) have shown that an injected sub-Keplerian flow can distribute the angular momentum and produce and sustain a Keplerian disk when viscosity is higher than a critical value, giving rise to the CT95 configuration of two component advective flow. However, this conclusion was drawn using simulations with the upper quadrant only. The conclusions drawn in the present paper with serve as the basis of of the next work with viscosity and radiative cooling. Most importantly, it would be clear if the resulting standard disk also exhibit such an oscillation. This will be discussed elsewhere.

{}

\end{document}